\documentclass[lettersize,journal]{IEEEtran}
\usepackage{amsmath,amsfonts}
\usepackage{algorithm}
\usepackage{array}
\usepackage[caption=false,font=normalsize,labelfont=sf,textfont=sf]{subfig}
\usepackage{textcomp}
\usepackage{stfloats}
\usepackage{url}
\usepackage{verbatim}
\usepackage{graphicx}

\usepackage{color}
\usepackage{algpseudocode}
\usepackage{xcolor}
\begin{document}
\bstctlcite{IEEEexample:BSTcontrol}
\title{Invertible Diffusion for Low-Memory Channel Gain Map Construction in Wireless Communication Networks}

\author{
Ruifeng Gao, {\em Member, IEEE},
Sen Li,
Jue Wang, {\em Member, IEEE},
Qiuming Zhu, {\em Senior Member, IEEE},\\
and Shu Sun, {\em Senior Member, IEEE}


\thanks{Ruifeng Gao and Sen Li are with the School of Transportation and Civil Engineering, Nantong University, Nantong 226019, China (grf@ntu.edu.cn; 2330310018@stmail.ntu.edu.cn).} \thanks{Jue Wang is with the School of Information Science and Technology, Nantong University, Nantong 226019, China (wangjue@ntu.edu.cn).} \thanks{Qiuming Zhu is with the College of Electronic and Information Engineering, Nanjing University of Aeronautics and Astronautics, Nanjing 211106, China (zhuqiuming@nuaa.edu.cn).} \thanks{Shu Sun is with the School of Information Science and Electronic Engineering, Shanghai Jiao Tong University, Shanghai 200240, China (shusun@sjtu.edu.cn).}

}

\markboth{Journal of \LaTeX\ Class Files,~Vol.~xx, No.~xx, xxx}%
{Shell \MakeLowercase{\textit{et al.}}: A Sample Article Using IEEEtran.cls for IEEE Journals}


\maketitle

\begin{abstract}

Channel gain maps (CGMs) enable propagation-aware services in edge-intelligent wireless communication networks, while diffusion-based CGM construction is memory intensive for on-device training or adaptation. This letter proposes InvDiff-CGM, an invertible diffusion framework that constructs CGMs from sparse measurements and environmental priors. By adopting invertible architectures in both the diffusion process and the U-Net noise estimator, InvDiff-CGM achieves near-constant training memory consumption. A prior-informed multi-scale injector further integrates environmental priors with sparse measurements to improve physical consistency and detail preservation. Experiments on RadioMap3DSeer show about an 85\% reduction in peak training memory and a PSNR of 38.02~dB, outperforming representative recent baselines. This validates the practicality of InvDiff-CGM for high-fidelity CGM construction under edge resource constraints.

\end{abstract}

\section{Introduction}
In edge-intelligent 6G networks, communication and sensing functions increasingly rely on propagation awareness under tight memory/compute budgets \cite{ref1, ref2}. Channel gain maps (CGMs) provide a spatial characterization of large-scale channel variation and are instrumental for coverage prediction, link adaptation, and environment-aware resource allocation \cite{ref3}.

CGMs can be obtained by ray tracing, spatial interpolation, or deep learning-based construction \cite{10891198}. In particular, ray tracing is physically interpretable but computationally intensive, which makes it difficult to satisfy real-time constraints at the edge \cite{ref4}. Spatial statistics-based techniques typically have lower computational cost, but their construction accuracy degrades in complex, highly non-linear electromagnetic environments \cite{ref5}. Among deep learning approaches, \cite{ref6} proposed RadioUNet, which leverages the encoder-decoder architecture to learn the non-linear mapping from environment to CGMs. While effective for function approximation, the convolutional operations in RadioUNet tend to smooth results, losing high-frequency details. The work \cite{ref7} introduced the RME-GAN framework based on generative adversarial networks (GANs), but GAN training may be unstable and prone to mode collapse, complicating CGM generation \cite{WGAN}.

Recently, diffusion models have shown promise in CGM construction due to their ability to capture fine-grained propagation transitions. A conditional diffusion framework was introduced in \cite{ref8} to recover CGMs from sparse observations, and RadioDiff \cite{ref9} further improved reconstruction by incorporating an adaptive filtering module. RadioDiff-$k^2$ embedded the Helmholtz equation as a physics-informed constraint to enhance physical consistency and generalization \cite{K2}. To reduce sampling latency, RadioDiff-Flux proposed trajectory midpoint reuse to improve inference efficiency while maintaining fidelity \cite{Flux}. Without environmental priors, RadioDiff-Inverse formulated CGM reconstruction as a Bayesian inverse problem, enabling training-free joint mapping and sensing from sparse measurements \cite{Inv}. RadioDiff-3D extended diffusion-based mapping to volumetric spaces and validated its effectiveness in complex air--ground environments \cite{3D}.


However, despite these advancements, most diffusion-based methods still rely on standard deep U-Net backbones. In particular, online learning and fine-tuning require caching intermediate activations across network layers and diffusion steps, leading to substantial training-time memory that scales approximately linearly with the unrolled depth, which makes deployment on resource-constrained edge nodes challenging.

To address these challenges, we propose \textbf{InvDiff-CGM} (\underline{\textbf{Inv}}ertible \underline{\textbf{Diff}}usion for \underline{\textbf{C}}hannel \underline{\textbf{G}}ain \underline{\textbf{M}}aps), a low-memory CGM reconstruction framework. 
InvDiff-CGM makes both the diffusion training pipeline and the U-Net noise estimator invertible, enabling on-demand activation reconstruction and near-constant peak memory. It further introduces a prior-informed multi-scale injector that fuses environmental geometry priors, such as building layouts, with sparse measurements to enhance physical plausibility, including path loss and shadowing, while preserving sharp boundaries.

\vspace{-0.5 em}

\section{System Model and Problem Formulation}
\label{sec:system_model}

We consider an edge-intelligent wireless network where edge nodes construct CGMs to support environment-aware communications within their coverage. Here, a CGM is a path-loss map defined on an \(H\times W\) grid, where \(H\) and \(W\) denote the map height and width. The CGM captures large-scale fading, including path loss and shadowing. We adopt a \(1\times1\)~m\(^2\) resolution, which is fine enough to resolve building-induced shadow transitions while averaging out rapid small-scale fading fluctuations; thus, the path loss is assumed approximately constant within each cell. Accordingly, the CGM is represented by \(\mathbf{X}\in\mathbb{R}^{H\times W}\), where \(x_{i,j}\) denotes the path loss at cell \((i,j)\). The BS location is \(\mathbf{c}_s=(h,d_x,d_y)\), where \(h\) is the antenna height and \((d_x,d_y)\) are the horizontal coordinates. Building geometry is represented by \(\mathbf{C}_b\in\mathbb{R}^{H\times W}\), where \(c^b_{i,j}=(\mathbf{C}_b)_{i,j}\) denotes the building height at \((i,j)\) and \(c^b_{i,j}=0\) indicates no building. For dense urban scenarios, we assume a flat terrain profile and focus on building-dominated propagation effects such as occlusion and reflection. The physical environment is summarized as \(\mathbf{C}=(\mathbf{c}_s,\mathbf{C}_b)\).


We aim to construct the CGM from the physical environment \(\mathbf{C}\) and sparse measurements  \(\mathbf{Y}\). Specifically, we train a neural network $\mu_{\theta}$ with parameters $\theta$ to estimate a high-accuracy map $\mathbf{\hat{X}} \in \mathbb{R}^{H \times W}$ from limited observations by leveraging environmental $\mathbf{C}$ to regularize the solution space. The network is trained by minimizing reconstruction loss $\mathcal{L}(\hat{X}, X)$. The CGM construction problem is formulated as 
\begin{equation}
\min_{\theta} \quad  \mathcal{L}(\mathbf{\hat{X}}, \mathbf{X}) \quad\quad \text{s.t.} \quad  \mathbf{\hat{X}} = \mu_{\theta}(\mathbf{Y}, \mathbf{C}), \mathbf{Y} = \mathcal{A}(\mathbf{X}) ,
    \label{eq:problem_formulation}
\vspace{-0.5 em}
\end{equation}
where $\mathbf{Y}$ denotes sparse measurements obtained via a random sampling operator $\mathcal{A}$. Regarding the practical acquisition of sparse geo-tagged received-power samples, standardized mechanisms such as Minimization of Drive Tests in 3GPP TS 37.320 \cite{3gpp_mdt} and the high device-density targets in ITU IMT-2020 \cite{itu_imt} support the feasibility of collecting such measurements in operational networks. These measurements enforce data consistency in the constructed CGM. In addition, environmental information $\mathbf{C}$ is incorporated as prior knowledge to restrict the feasible solution space, encouraging constructions that are consistent with the underlying propagation physics.


To solve this problem, we propose a cascaded deep generative network based on diffusion models that explicitly parametrizes the iterative denoising process. Following the denoising diffusion null-space model \cite{ref11}, we formulate each denoising step as a deterministic denoising diffusion implicit model (DDIM) update \cite{ref12} consisting of three sub-steps, noise prediction, data-consistency projection, and state update. This design yields a differentiable and deterministic mapping from sparse observations to the constructed CGM. 

Specifically, we unroll the $T$-step denoising process from the initial state $\mathbf{X}_{T}$ to the target state $\mathbf{X}_{0}$ as a cascaded neural network $\mathcal{F} = \mathcal{F}_{1} \circ \cdots  \mathcal{F}_{t} \cdots \circ\mathcal{F}_{T}$, where $\circ$ denotes function composition. Note that uppercase $T$ represents the maximum number of diffusion steps, whereas lowercase $t \in \{0,1, \dots, T\}$ denotes the current step index. $\mathcal{F}_{t}$ is the single-step update operator that performs noise prediction and data-consistency enforcement at the iteration $t$. However, during training, back-propagation through the unrolled iterations requires storing intermediate variables, resulting in substantial GPU memory overhead that scales with the number of iteration steps $T$. To address this, we propose InvDiff-CGM, which reduces training-time memory consumption to a near-constant level while preserving construction accuracy, thereby enabling high-quality CGM construction on edge nodes.

\begin{figure}[!ht]
\centering
\includegraphics[width=3.5in]{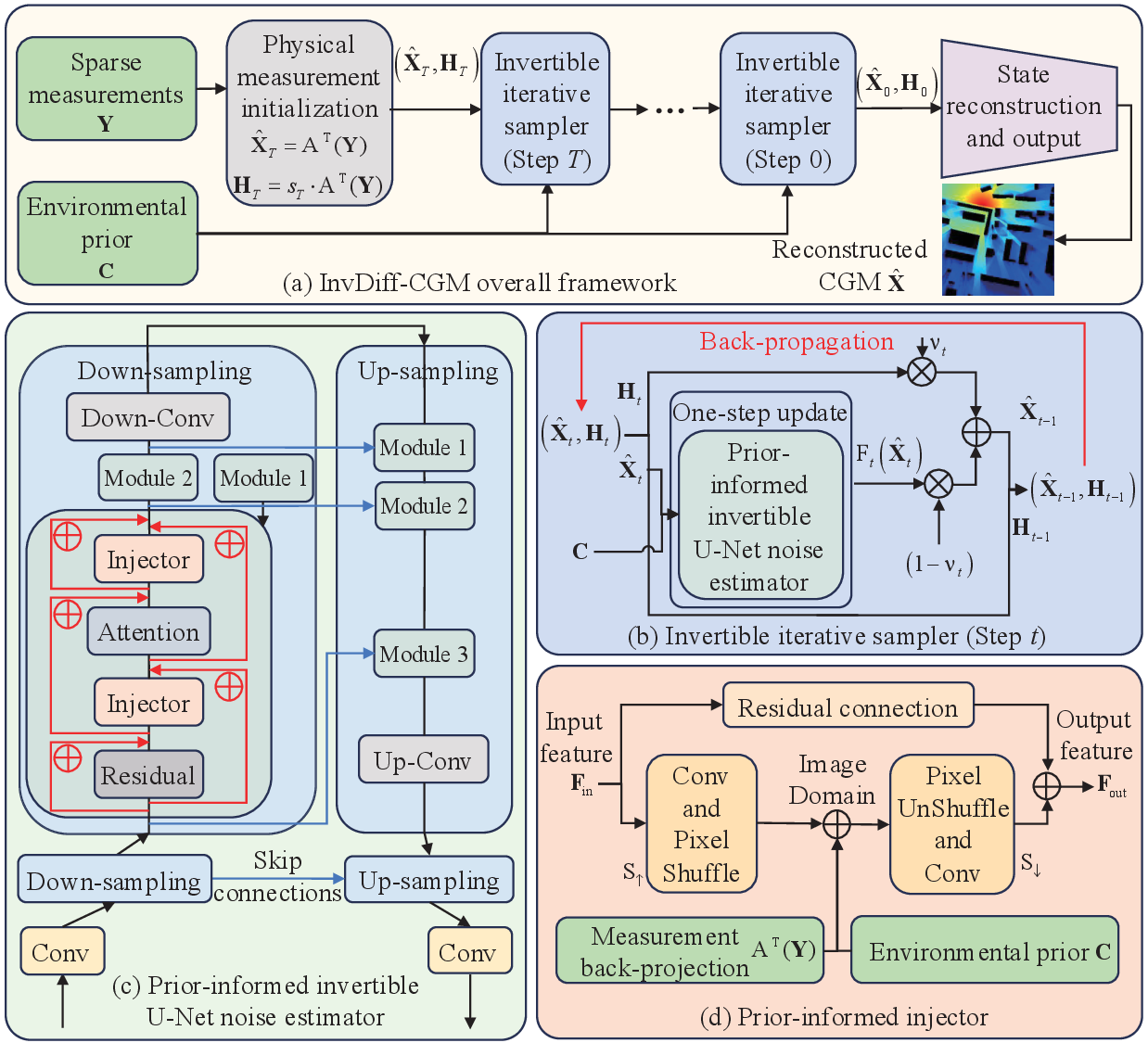}
\caption{Overview of InvDiff-CGM. }
\label{fig1}
\vspace{-1.5 em}
\end{figure}

\section{Proposed InvDiff-CGM Framework}
\label{sec:method}

\subsection{Overall Framework}
We propose InvDiff-CGM to solve \eqref{eq:problem_formulation}. As shown in Fig.~\ref{fig1}(a), it recasts the multi-step diffusion sampling procedure as a low-memory deterministic solver for the underlying inverse problem, consisting of three stages.

\subsubsection{Physical Measurement Initialization} To accelerate convergence and incorporate measurement priors, we initialize the iteration using the back-projection of the sparse measurements. Specifically, the initial state is set to $\mathbf{\hat{X}}_{T} = \mathcal{A}^{\mathrm{T}}(\mathbf{Y})$, where $\mathcal{A}^{\mathrm{T}}$ denotes the adjoint of the sampling operator $\mathcal{A}$. To accommodate the dual-channel input of the invertible network and maintain dimensional consistency, we introduce a learnable scalar $s_{T}$ and an auxiliary variable $\mathbf{H}_T\in \mathbb{R}^{H \times W}$ is initialized as $\mathbf{H}_{T} = s_{T} \cdot \mathcal{A}^{\mathrm{T}}(\mathbf{Y})$, with $s_{T}$ initialized to 1. This initialization ensures the iteration is physically consistent with constraints and provides a strong prior for subsequent updates.

\subsubsection{Invertible Denoising Iteration}
This stage progressively recovers high-frequency details using $T$ cascaded networks. In conventional unrolled diffusion solvers, a major problem is the need to store intermediate variables at every iteration for back-propagation, leading to significant memory consumption. To address this, we adopt an invertible connection pattern and reformulate the reverse diffusion update from step $t$ to $t-1$ as a mathematically invertible mapping. Specifically, each update operator $\mathcal{F}_t$ integrates noise prediction, data-consistency correction, and state update. Given the current state $(\mathbf{\hat{X}}_t, \mathbf{H}_t)$, the model computes the next denoised state $(\mathbf{\hat{X}}_{t-1}, \mathbf{H}_{t-1})$, which is closer to the target distribution, using the forward rule of the invertible transform. During back-propagation, the required inputs can be reconstructed on the fly from the outputs, thereby avoiding variable caching and substantially reducing memory consumption.

\subsubsection{State Reconstruction and Output}
When $t=0$, we obtain the final state $(\mathbf{\hat{X}}_0, \mathbf{H}_0)$. The construction is given by $\mathbf{X}_0=\mathbf{\hat{X}}_0+s_0 \mathbf{H}_0$, where $s_0$ is initialized to 1. $s_0$ adaptively fuses the main feature $\mathbf{\hat{X}}_0$ with the auxiliary $\mathbf{H}_0$ within the invertible architecture, preserving the information carried by input channels and maps it back to the original signal domain.

In InvDiff-CGM, we adopt a dual invertible design comprising a multi-step invertible iterative sampler and a prior-informed invertible U-Net noise estimator. The sampler updates the state at each iteration, while the invertible U-Net performs denoising, improving both efficiency and reconstruction accuracy.

\vspace{-0.5 em}

\subsection{Multi-step Invertible Iterative Sampler}

\subsubsection{Invertible Iterative Update Unit}
To address the challenge of limited memory of edge devices, we propose the multi-step invertible iterative sampler shown in Fig.~\ref{fig1}(b), where Step \(t\) denotes an arbitrary iteration index. By introducing $\mathbf{H}_t$ and an invertible coupling pattern \cite{ref13}, each sampling step is reformulated as an invertible transformation. During back-propagation, intermediate variables are recomputed on demand from the outputs, thereby substantially reducing training-time memory. Specifically, at iteration $t$ we represent the state as a dual-channel tuple $(\mathbf{\hat{X}}_{t}, \mathbf{H}_{t})$. The forward update is defined by the invertible coupling rule
\vspace{-0.2 em}
\begin{equation}
        \mathbf{\hat{X}}_{t-1} = (1 - v_{t}) \times \mathcal{F}_{t}(\mathbf{\hat{X}}_{t}) + v_{t} \times \mathbf{H}_{t},\quad
        \mathbf{H}_{t-1} = \mathbf{\hat{X}}_{t},
    \label{eq:invertible_update}
\end{equation}
where $v_{t}$ is a learnable scalar that adaptively balances the single-step operator $\mathcal{F}_{t}(\cdot)$ and $\mathbf{H}_{t}$. Because the second assignment $\mathbf{H}_{t-1} = \mathbf{\hat{X}}_{t}$ preserves $\mathbf{\hat{X}}_{t}$ exactly, the entire transformation is strictly mathematically invertible. Its inverse follows directly from (\ref{eq:invertible_update}). As a result, during the forward pass, we only store the terminal state $(\mathbf{\hat{X}}_{0}, \mathbf{H}_{0})$. The input state $\lambda(\mathbf{\hat{X}}_{t}, \mathbf{H}_{t})$ at each step and the internal variables are reconstructed on demand by applying the inverse mapping, and freed immediately after the gradients are computed.

\subsubsection{One-Step Update with Noise Prediction and Data Consistency Projection} In (\ref{eq:invertible_update}), $\mathcal{F}_{t}(\mathbf{\hat{X}}_{t})$ denotes the 
one-step transition that maps the noisy state $\mathbf{\hat{X}}_{t}$ at iteration $t$ to the denoised estimate $\mathbf{X}_{t-1}^{\text{diff}}$ for the next iteration. This update jointly incorporates the diffusion prior and the physical measurement constraints, and consists of the following three sub-steps.

(1) Initial Estimation: We first estimate the noise $\hat{e}_{t}$ using the prior-informed invertible U-Net noise estimator $\epsilon_{\Theta}$, 
\begin{equation}
    \hat{e}_{t} = \epsilon_{\Theta}(\text{concat}(\mathbf{\hat{X}}_{t}, \mathbf{C}), t),
    \label{eq:noise_prediction}
\end{equation}
where $\text{concat}(\cdot)$ denotes the concatenation for injecting the environmental prior $\mathbf{C}$. Following the DDIM sampling rule \cite{ref12}, we then obtain an unconstrained estimate $\mathbf{\hat{X}}_{0|t}$ of the noise-free map $\mathbf{X}_{0}$ from $\mathbf{\hat{X}}_{t}$ and $\hat{e}_{t}$,
\begin{equation}
    \mathbf{\hat{X}}_{0|t} = \frac{1}{\sqrt{\bar{\alpha}_{t}}}(\mathbf{\hat{X}}_{t} - \sqrt{1 - \bar{\alpha}_{t}} \cdot \hat{e}_{t}),
    \label{eq:x0_prediction}
\end{equation}
where $\bar{\alpha}_{t} = \prod_{i=0}^{t} \alpha_{i}$  with $\alpha_{0}=1$ denotes the cumulative noise schedule. The coefficient $\bar{\alpha}_{t}$ controls the signal-to-noise ratio at step $t$, and $\mathbf{\hat{X}}_{0|t}$ is the model’s maximum-likelihood estimate under the diffusion prior at the current iteration.

(2) Data Consistency Correction: Although the unconstrained estimate $\mathbf{\hat{X}}_{0|t}$ captures the features of the real CGM, it does not necessarily satisfy the measurement constraint. We therefore enforce data consistency by projecting $\mathbf{\hat{X}}_{0|t}$ onto the feasible set defined by the measurements,
\begin{equation}
    \mathbf{\overline{X}_{0|t}} = \mathbf{\hat{X}}_{0|t} - \eta \cdot \mathcal{A}^{T}(\mathcal{A}(\mathbf{\hat{X}}_{0|t}) - \mathbf{Y}).
    \label{eq:dc_correction}
\end{equation}
Here, $\mathcal{A}$ is an orthogonal sampling operator satisfying $\mathcal{A}\mathcal{A}^{\mathrm{T}} = \mathbf{I}$, and we have $\mathcal{A}^{\mathrm{T}} = \mathcal{A}^{\dagger}$. By setting $\eta = 1$, \eqref{eq:dc_correction} performs an orthogonal projection onto the affine subspace $\mathbf{Y}=\mathcal{A}(\mathbf{X})$, yielding the solution $\mathbf{\overline{X}_{0|t}} = \mathcal{A}^{\dagger}\mathbf{Y} + (\mathbf{I} - \mathcal{A}^{\dagger}\mathcal{A})\mathbf{\hat{X}}_{0|t}$. This projection enforces strict data consistency without the iterative step-size tuning required by the traditional schemes.

(3) State Update: Finally, following the deterministic DDIM update rule, we combine the data-consistent estimate $\mathbf{\overline{X}_{0|t}}$ with the predicted noise $\hat{e}_{t}$ to obtain the next state $\mathbf{X}_{t-1}^{\text{diff}}$,
\begin{equation}
    \mathbf{X}_{t-1}^{\text{diff}} = \sqrt{\bar{\alpha}_{t-1}} \times \mathbf{\overline{X}_{0|t}} + \sqrt{1 - \bar{\alpha}_{t-1}} \cdot \hat{e}_{t}.
    \label{eq:state_update}
\end{equation}

The above three sub-steps are encapsulated in $\mathcal{F}_{t}(\mathbf{\hat{X}}_{t})$. Substituting $\mathcal{F}_{t}(\mathbf{\hat{X}}_{t})$ into (\ref{eq:invertible_update}) preserves the invertibility, leading to the low-memory consumption of the overall iterative process.

\subsection{Prior-Informed Invertible U-Net Noise Estimator}
\subsubsection{Invertible U-Net Reconstruction and Memory Optimization}
While the multi-step invertible iterative sampler eliminates memory growth with the diffusion horizon, a standard deep U-Net still incurs substantial memory overhead due to the intermediate variables generated within each forward pass. To further reduce the memory consumption, we extend the invertible design to the U-Net itself and construct an invertible U-Net noise estimator. As shown in Fig.~\ref{fig1}(c), it is organized with two down-sampling and two up-sampling stages.

For memory optimization, the computationally intensive components in each stage—including residual blocks, attention blocks, and prior-informed injectors—are packaged into independent reversible modules (Modules 1–3 in Fig.~\ref{fig1}(c)). Each down-sampling stage comprises two reversible modules followed by a strided convolutional layer, whereas each up-sampling stage comprises three reversible modules followed by a transposed-convolution layer. Within each reversible module, the invertible coupling pattern (red connections in Fig.~\ref{fig1}(c)) splits the input into two feature streams that are updated in an alternating manner. As a result, during back-propagation, inputs and internal variables are reconstructed on demand from the outputs, avoiding activation caching in the forward pass and substantially reducing memory.

For inter-module operations that are not strictly invertible, most notably resolution changing layers and skip connections (blue connections in Fig.~\ref{fig1}(c)), we retain the standard U-Net design. By confining most computation to reversible modules, the proposed architecture markedly reduces the memory consumption, thereby enabling large-scale end-to-end training on a single consumer-grade GPU of the edge nodes.


\subsubsection{Multi-Scale Prior-Informed Injector}
To ensure that the generated representations remain consistent with sparse observations after measurement projection, we embed prior-informed injector modules along the multi-scale down-sampling and up-sampling paths of the U-Net to mitigate the lack of physical constraints. As shown in Fig.~\ref{fig1}(d), given a feature map $\mathbf{F_{in}} \in \mathbb{R}^{(H/r)\times(W/r)\times C}$ at a U-Net layer, where $r$ is the scaler factor of the current layer relative to the original resolution, the injector first lifts $\mathbf{F_{in}}$ to the high-resolution image domain. It then fuses this representation with the measurement back-projection $\mathcal{A}^{T}(\mathbf{Y})$ and the environmental prior through multi-channel concatenation. Finally, the fused tensor is mapped back to the feature domain to produce a residual correction. This procedure is formulated as
\begin{equation}
\begin{aligned}
    \mathbf{F_{out}}&= \mathbf{F_{in}} \\&+ \mathcal{S}_{\downarrow}(\text{Conv}_{2}(\text{Concat}(\text{Conv}_{1}(\mathcal{S}_{\uparrow}(\mathbf{F_{in}})), \mathcal{A}^{T}(\mathbf{Y}), \mathbf{C}))).
    \label{eq:injector}     
\end{aligned}
\end{equation}
where $\mathcal{S}_{\uparrow}$ and $\mathcal{S}_{\downarrow}$ denote PixelShuffle and PixelUnshuffle operations, respectively. $\text{Conv}_{1}$ is a $1 \times 1$ convolution that compresses the upsampled features into a single channel to align with the physical inputs, while $\text{Conv}_{2}$ maps the fused multi-channel tensor back to the feature-space dimension. 

By applying this injection at multiple resolutions within the U-Net, sparse measurements and environmental priors are integrated throughout the feature hierarchy, which enhances high-frequency detail recovery and enforces physical consistency, thereby improving CGM construction quality.


\section{Performance Evaluation}
\label{sec:experiments}

We evaluate InvDiff-CGM on the public RadioMap3DSeer dataset \cite{ref14} with a 3.5~GHz carrier frequency, where building heights range from 6.6~m to 19.8~m and the transmitter is deployed 3~m above a rooftop higher than 16.5~m. Each CGM is rasterized into a \(256\times256\) image, with \(1~\text{m}^2\) per pixel. InvDiff-CGM uses the U-Net from pre-trained Stable Diffusion v1.5 as the backbone and is trained with Adam (initial learning rate \(1\times10^{-4}\)) and a MultiStepLR scheduler (milestones at epochs 25 and 38). We use an \(L_1\) loss, set the diffusion steps to \(T=3\), and enable automatic mixed precision. All experiments are conducted on a single NVIDIA RTX~4090 GPU (48~GB). We compare against six baselines: RadioUNet \cite{ref6}, RME-GAN \cite{ref7}, RadioDiff \cite{ref9}, RadioDiff-\(k^2\) \cite{K2}, RadioDiff-Flux \cite{Flux}, and RadioDiff-Inverse \cite{Inv}, using the same train/test split and the original architectures and hyperparameters. 
We evaluate construction quality using peak signal-to-noise ratio (PSNR), structural similarity index measure (SSIM), normalized mean squared error (NMSE), and root mean squared error (RMSE).
Let ${\rm MSE}=\|\hat{\mathbf{X}}-\mathbf{X}\|_F^2/(HW)$, then 
${\rm PSNR}=10\log_{10}\!\big(X_{\max}^2/{\rm MSE}\big)$ (fidelity in dB), where \(X_{\max}\) denotes the maximum possible (or observed) path-loss value used in computing PSNR, 
${\rm SSIM} \in[0,1]$ measures structural similarity, 
${\rm NMSE}=\|\hat{\mathbf{X}}-\mathbf{X}\|_F^2/\|\mathbf{X}\|_F^2$, and 
${\rm RMSE}=\sqrt{{\rm MSE}}$.

\begin{table}[!t]
\caption{Quantitative comparison on the RadioMap3DSeer test set}
\label{tab:quantitative_comparison}
\centering
\begin{tabular}{l c c c c}
\hline
\textbf{Method} & \textbf{PSNR (dB)$\uparrow$} & \textbf{SSIM$\uparrow$} & \textbf{NMSE$\downarrow$} & \textbf{RMSE$\downarrow$} \\
\hline
RadioUNet  & 28.79 & 0.8671 & 0.01308 & 0.0318 \\
RME-GAN  & 23.68 & 0.7289 & 0.12487 & 0.0982 \\
RadioDiff  & 35.52 & 0.9435 & 0.00410 & 0.0191 \\
RadioDiff-$k^2$  & 36.19 & 0.9461 & 0.00274 & 0.0161 \\
RadioDiff-Inverse  & 33.23 & 0.9159 & 0.00519 & 0.0210 \\
RadioDiff-Flux  & 35.37 & 0.9400 & 0.00329 & 0.0177 \\
\textbf{Ours} & \textbf{ 38.02}$^{\mathrm{*}}$ & \textbf{ 0.9634}$^{\mathrm{*}}$ & \textbf{ 0.00196}$^{\mathrm{*}}$ & \textbf{ 0.0123}$^{\mathrm{*}}$ \\
\hline
\multicolumn{5}{l}{\footnotesize $^{\ast}$Best result among all methods.}
\end{tabular}
\vspace{-1 em}
\end{table}

\begin{figure}[!t]
\centering
\includegraphics[width=2.5 in]{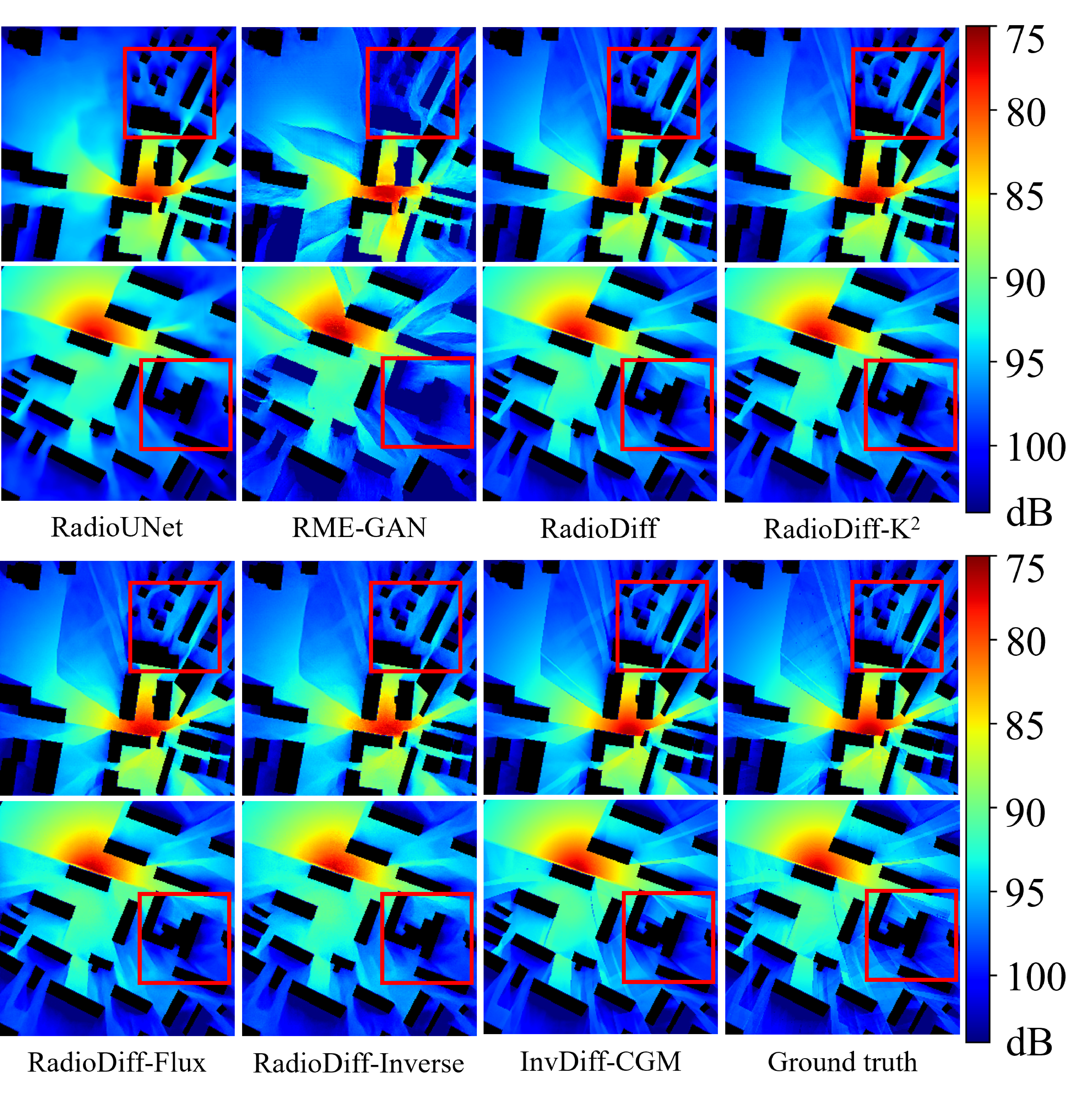}
\vspace{-1 em}
\caption{Qualitative comparison of constructed CGMs on representative RadioMap3DSeer test scenes.}
\label{fig2}
\vspace{-1 em}
\end{figure}

\subsection{Performance of InvDiff-CGM}
 Table~\ref{tab:quantitative_comparison} reports the quantitative results on the RadioMap3DSeer test set. InvDiff-CGM achieves the best overall performance across PSNR, SSIM, NMSE, and RMSE, attaining 38.02~dB PSNR and 0.9634 SSIM. It consistently outperforms RadioUNet, RME-GAN, RadioDiff, and recent RadioDiff variants. In particular, compared with RadioDiff-\(k^2\), InvDiff-CGM provides a 1.83~dB PSNR gain, suggesting that explicitly injecting multi-scale geometric priors is more effective in preserving sharp building-induced transitions. Compared with RadioDiff-Inverse, InvDiff-CGM benefits from environmental priors and yields substantially higher fidelity in shadowed regions. InvDiff-CGM also outperforms RadioDiff-Flux, indicating that improved sampling efficiency does not necessarily translate into higher reconstruction fidelity under the same evaluation protocol.
Fig.~\ref{fig2} further corroborates these trends. InvDiff-CGM produces reconstructions closest to the ground truth, with clearer building boundaries and more faithful abrupt attenuation transitions caused by occlusion and reflection. Notably, these gains are achieved using a single NVIDIA RTX~4090 GPU, whereas the baseline four RadioDiff methods training relies on a \(4\times\)A100 cluster, highlighting a more favorable option for edge-intelligent deployments.

\subsection{Ablation Study}
\begin{table}[!t]
\caption{Comparison of Peak GPU Memory Utilization During Training}
\label{tab:memory_comparison}
\centering
\begin{tabular}{c c c c}
\hline
\textbf{$T$} & \textbf{Invertible (GB)} & \textbf{Non-Invertible (GB)} & \textbf{Memory Reduction} \\
\hline
1 & 5.2 & 16.7 & 68.69\% \\
2 & 6.9 & 33.3 & 79.30\% \\
3 & 6.9 & 49.0 & 85.59\% \\
\hline
\end{tabular}
\vspace{-1 em}
\end{table}

\subsubsection{Ablation Study on Invertible Architecture}
Table~\ref{tab:memory_comparison} compares the peak training memory with and without the invertible modules. Without invertibility, standard back-propagation must cache intermediate variables at each diffusion step, causing memory consumption to scale linearly with the number of steps $T$, i.e., $\mathcal{O}(T)$. As $T$ increases, the peak memory grows from 16.7 GB at $T=1$ to 49 GB at $T=3$, exceeding the capacity of an NVIDIA RTX 4090 and making end-to-end training infeasible. In contrast, the invertible design keeps near-constant memory complexity, approximately $\mathcal{O}(1)$.


\subsubsection{Ablation Study on the Prior-Informed Injector}
\begin{figure}[!t]
\centering
\includegraphics[width=3.5 in]{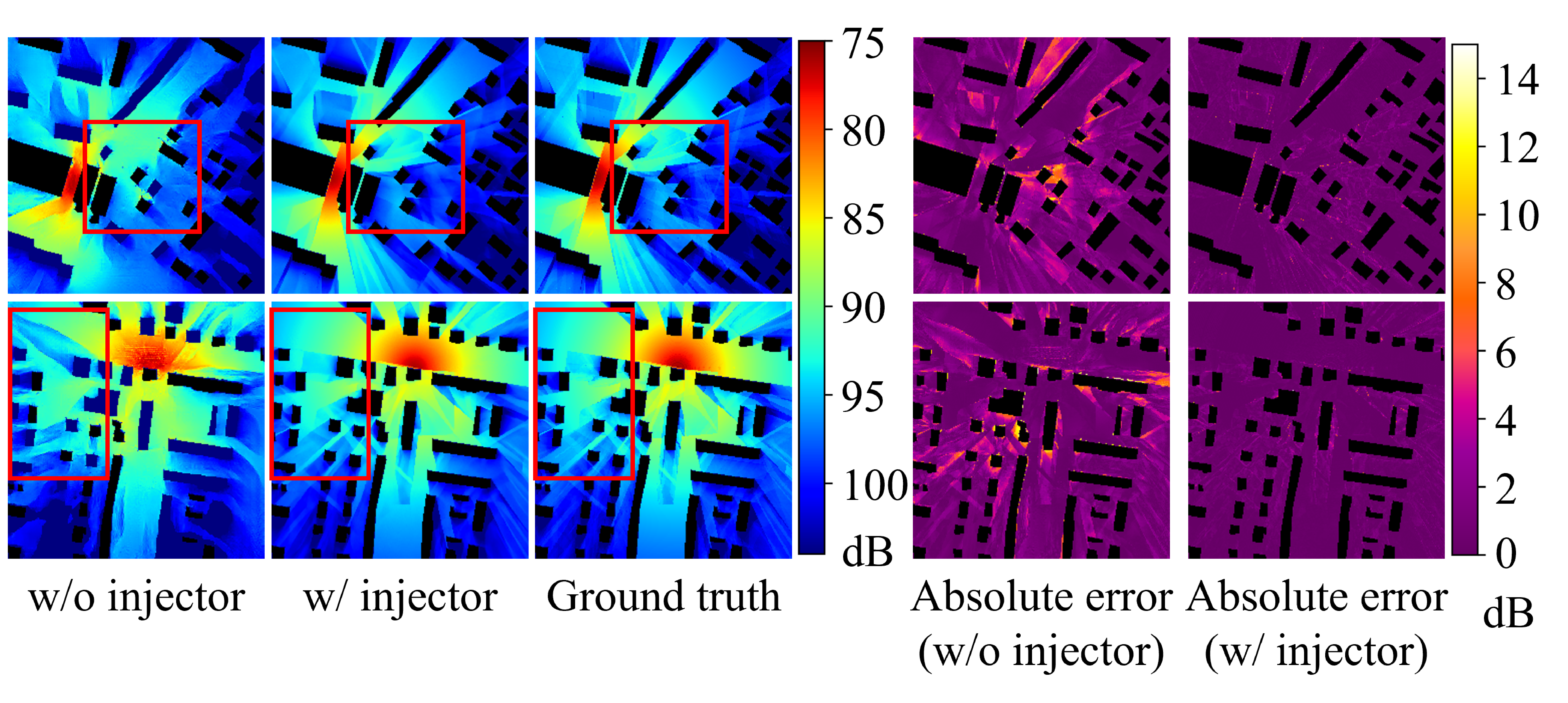}
\vspace{-1.5 em}
\caption{Comparison of InvDiff-CGM w/ and w/o the prior-informed injector.}
\label{fig3}
\vspace{-1.5 em}
\end{figure}

To quantify the contribution of the prior-informed injector, we remove the injector modules from all U-Net layers while keeping the remaining settings unchanged. Only a global data-consistency projection is applied at the end of the diffusion process. As shown in Fig.~\ref{fig3}, disabling the injectors leads to pronounced blurriness and distortion in regions with abrupt signal variations, and building-induced occlusion boundaries become less distinct. 
In contrast, the proposed injector continuously integrates sparse measurement constraints and environmental priors into multi-scale U-Net features throughout CGM generation, thereby improving the overall construction quality.

\subsection{Runtime--Memory Trade-off and Limitations}
\label{sec:discussion}
We analyze the trade-off between memory savings and computational overhead. During inference, the invertible design performs a standard forward pass and achieves a latency of 0.15 s per map, identical to the non-invertible baseline, indicating no additional cost for online deployment. During training, reconstructing intermediate states increases the per-epoch runtime by approximately 38\%, effectively converting a hard memory constraint into a manageable runtime overhead. To explicitly quantify this per-iteration overhead, we profiled both models with batch size of 2. Although the non-invertible baseline is theoretically faster per iteration (1.23 s), our InvDiff-CGM takes 1.69 s. We further validate the practical necessity of this trade-off via an online adaptation experiment: when fine-tuning on 50 unseen maps (with a target batch size of 4), the non-invertible baseline encounters an out-of-memory (OOM) failure under the same hardware setting. In contrast, InvDiff-CGM successfully bypasses the memory bottleneck, completing the adaptation in 263.7 s and improving the PSNR to 38.19 dB. These results indicate that the proposed invertible architecture substantially lowers the hardware barrier for diffusion-based CGM learning under edge resource constraints. We also note two limitations: recomputation increases training latency for time-sensitive adaptation, and the flat-terrain assumption may limit accuracy in mountainous environments. Terrain-aware modeling and more memory-efficient attention variants are promising extensions.

\section{Conclusion}
\label{sec:conclusion}
This letter proposes InvDiff-CGM, an invertible diffusion framework for CGM construction from sparse measurements and environmental priors. Invertible designs in both the diffusion process and U-Net noise estimator enable near-constant training memory, while a multi-scale prior-informed injector improves physical consistency and detail fidelity. Experiments show that InvDiff-CGM outperforms baselines and yields constructions closer to the ground truth, supporting its use in edge-intelligent networks. Future work will model time-varying vehicular shadowing in urban scenarios by extending the current spatial model to a spatio-temporal generative framework, where real-time traffic states are introduced as additional conditioning to enable dynamic CGM updates.



\bibliography{ref}

\end{document}